\documentclass[conference]{IEEEtran}
\IEEEoverridecommandlockouts
\usepackage{cite}
\usepackage{amsmath,amssymb,amsfonts}
\usepackage{graphicx}
\usepackage{textcomp}
\usepackage[prologue,usenames,dvipsnames,svgnames,table]{xcolor}
\usepackage{epsfig,endnotes}
\usepackage{booktabs} 
\usepackage{multirow}
\usepackage{tipa}
\usepackage{subcaption} 
\usepackage{pgfplots}
\usepackage{pdfcomment}
\usepackage{rotating}
\usepackage{lscape}
\usepackage{lipsum}
\usepackage{wrapfig}  
\usepackage{balance}
\usepackage{slashbox}
\usepackage{amsmath}
\usepackage{algorithm}
\usepackage[noend]{algpseudocode}
\usepackage{cleveref}
\usepackage{nameref}
\crefformat{section}{\S#2#1#3} 
\crefformat{subsection}{\S#2#1#3}
\crefformat{subsubsection}{\S#2#1#3}
\usepackage{tcolorbox}
\usepackage[inline]{enumitem} 
\usepackage{amssymb}
\usepackage{pifont}
\usepackage{array}
\usepackage{tabularx}

\newcommand\clearrow{\global\let\rowmac\relax}
\clearrow
\usepackage{hyperref}
\usepackage[xindy,toc]{glossaries}

\newcommand{\RQone}{How does \VT{}'s dynamicity impact the performance of threshold-based labeling strategies?}

\newcommand{\RQtwo}{Which labeling strategy should be used to label apps based on their \VT{} scan reports accurately?}

\newcommand{\RQthree}{What are the limitations of \texttt{VirusTotal} and how can they be addressed?}

\newcommand{\Eleda}{\emph{Eleda}}
\newcommand{\VT}{\texttt{VirusTotal}}

\newcommand{\latestDate}{September 27$^{th}$, 2019}
\newcommand{\futureDate}{November 8$^{th}$, 2019}

\newtcolorbox{mydef}{title=Definition}

\definecolor[named]{ACMBlue}{cmyk}{1,0.1,0,0.1} 
\definecolor[named]{ACMYellow}{cmyk}{0,0.16,1,0} 
\definecolor[named]{ACMOrange}{cmyk}{0,0.42,1,0.01} 
\definecolor[named]{ACMRed}{cmyk}{0,0.90,0.86,0} 
\definecolor[named]{ACMLightBlue}{cmyk}{0.49,0.01,0,0} 
\definecolor[named]{ACMGreen}{cmyk}{0.20,0,1,0.19} 
\definecolor[named]{ACMPurple}{cmyk}{0.55,1,0,0.15} 
\definecolor[named]{ACMDarkBlue}{cmyk}{1,0.58,0,0.21} 

\newcounter{mylabelcounter}
\newcommand{\labelText}[2]{%
	#1\refstepcounter{mylabelcounter}%
	\immediate\write\@auxout{%
	  \string\newlabel{#2}{{1}{\thepage}{{\unexpanded{#1}}}{mylabelcounter.\number\value{mylabelcounter}}{}}%
	}%
}

\makeglossaries
\newacronym[plural=APIs]{api}{API}{Application Programming Interface}
\newacronym[plural=APKs]{apk}{APK}{Android Package}
\newacronym[plural=GFGs]{cfg}{CFG}{Control Flow Graph}
\newacronym[plural=DFGs]{dfg}{DFG}{Data Flow Graph}
\newacronym[plural=DDGs]{ddg}{DDG}{Data Depdendency Graph}
\newacronym{gui}{GUI}{Graphical User Interface}
\newacronym{io}{IO}{input-output}
\newacronym{loc}{LOC}{Lines of Code}
\newacronym[plural=IDEs]{ide}{IDE}{Integrated Development Environment}
\newacronym{jdk}{JDK}{Java Development Kit}
\newacronym[plural=SDKs]{sdk}{SDK}{Software Development Kit}
\newacronym{ndk}{NDK}{Native Development Kit}
\newacronym{xml}{XML}{Extensible Markup Language}
\newacronym{jar}{JAR}{Java ARchive}
\newacronym[plural=DLLs]{dll}{DLL}{Dynamic-link Library}
\newacronym{csv}{CSV}{Comma-Separated Values}
\newacronym{sql}{SQL}{Structured Query Language}
\newacronym{crud}{CRUD}{Create, Read, Update, and Delete}
\newacronym[plural=PUAs]{pua}{PUA}{Potentially Unwanted Application}
\newacronym{json}{JSON}{JavaScript Object Notation}
\newacronym{ssim}{SSIM}{Structural Similarity Index}

\newacronym{os}{OS}{Operating System}
\newacronym{dma}{DMA}{Direct Memory Access}
\newacronym[plural=AVDs]{avd}{AVD}{Android Virtual Device}
\newacronym[plural=VMs]{vm}{VM}{Virtual Machine}
\newacronym{gps}{GPS}{Global Positioning System}
\newacronym[plural=PLCs]{plc}{PLC}{Programmable Logic Controller}
\newacronym{sim}{SIM}{Subscriber Identification Module}
\newacronym{mac}{MAC}{Media Access Control}
\newacronym{imei}{IMEI}{International Mobile Equipment Identity}
\newacronym{dcl}{DCL}{Dynamic Code Loading}
\newacronym{sms}{SMS}{Short Message Service}
\newacronym[plural=URLs]{url}{URL}{Uniform Resource Locator}
\newacronym{art}{ART}{Android Runtime}
\newacronym[plural=IPs]{ip}{IP}{Internet Protocol}

\newacronym{ai}{AI}{Artificial Intelligence}
\newacronym{ml}{ML}{Machine Learning}
\newacronym{knn}{KNN}{K-Nearest Neighbors}
\newacronym[plural=SVMs]{svm}{SVM}{Support Vector Machine}
\newacronym[plural=SVCs]{svc}{SVC}{Linear Support Vector Machine}
\newacronym[plural=DTs]{dt}{DT}{Decision Tree}
\newacronym[plural=RFs]{rf}{RF}{Random Forest}
\newacronym{gnb}{GNB}{Gaussian Naive Bayes}
\newacronym[plural=HMMs]{hmm}{HMM}{Hidden Markov Model}
\newacronym{em}{EM}{Expectation Maximization}
\newacronym{pca}{PCA}{Principal Component Analysis}
\newacronym{mcc}{MCC}{Matthews Correlation Coefficient}

\glsenablehyper

\def\BibTeX{{\rm B\kern-.05em{\sc i\kern-.025em b}\kern-.08em
    T\kern-.1667em\lower.7ex\hbox{E}\kern-.125emX}}
\begin{document}

\title{Towards Accurate Labeling of Android Apps for Reliable Malware Detection}

\author{
	\IEEEauthorblockN{Aleieldin Salem}
	\IEEEauthorblockA{\textit{Technische Universit\"{a}t M\"{u}nchen}\\
	Garching bei M\"{u}nchen, Germany \\
	salem@in.tum.de}
}

\maketitle

\begin{abstract}
In training their newly-developed malware detection methods, researchers rely on threshold-based labeling strategies that interpret the scan reports provided by online platforms, such as \VT{}. 
The dynamicity of this platform renders those labeling strategies unsustainable over prolonged periods, which leads to inaccurate labels. 
Using inaccurately labeled apps to train and evaluate malware detection methods significantly undermines the reliability of their results, leading to either dismissing otherwise promising detection approaches or adopting intrinsically inadequate ones. 
The infeasibility of generating accurate labels via manual analysis and the lack of reliable alternatives force researchers to utilize \VT{} to label apps. 
In the paper, we tackle this issue in two manners. 
Firstly, we reveal the aspects of \VT{}'s dynamicity and how they impact threshold-based labeling strategies and provide actionable insights on how to use these labeling strategies given \VT{}'s dynamicity reliably. 
Secondly, we motivate the implementation of alternative platforms by (a) identifying \VT{} limitations that such platforms should avoid, and (b) proposing an architecture of how such platforms can be constructed to mitigate \VT{}'s limitations.
\end{abstract}


\pagestyle{plain}
\section{Introduction}
\label{sec:introduction}

The generation of reliable ground truths for malicious and benign applications (hereafter apps) is fundamental for implementing and evaluating effective malware detection methods. 
Assigning apps inaccurate labels (e.g., labeling malicious apps as benign) might impact the reliability of studies that inspect trends adopted by malicious apps, and, more importantly, might impede the development of effective detection methods \cite{hurier2017euphony,hurier2016lack,miller2016reviewer,kantchelian2015better}. 
Manual analysis and labeling of apps is arguably the most reliable method to label apps. 
However, it can neither cope with the frequent release of malware nor the requirement of some detection methods (e.g., \gls{ml}-based methods) of large numbers of labeled apps for training and validation. 
Consequently, researchers turn to online platforms, such as \VT{} \cite{virustotal2019}, to label apps in their datasets as malicious and benign. 

\VT{} does not label apps as malicious and benign.
Given the hash of an app or its executable, the platform provides the scan results from about 60 different commercial antiviral software \cite{wang2018beyond,li2017understanding,suarez2017droidsieve,wei2017deep,yang2017malware}.
So, it is up to the platform's user to decide upon strategies to interpret such information to label apps as malicious and benign. 
Unfortunately, there are no standard procedures for interpreting the scan results acquired from \VT{} to label apps, which leads researchers to use their intuitions and adopt ad hoc threshold-based strategies to label the apps in the datasets used to train and evaluate their detection methods.
In essence, threshold-based labeling strategies deem an app as malicious if the number of antiviral scanners labeling the apps as malicious meets a certain threshold. 
For example, based on \VT{}'s scan reports, Li et al.\ labeled the apps in their \emph{Piggybacking} dataset as malicious if at least one scanner labeled them as malicious  ~\cite{li2017understanding}, Pendlebury et al.\ labeled an app as malicious if four or more scanners did so ~\cite{pendlebury2019}, and 
Wei et al.\ labeled apps in the \emph{AMD} dataset as malicious if 50\% or more of the total scanners labeled an app as such ~\cite{wei2017deep}. 

Some of the aforementioned threshold-based labeling strategies may indeed accurately label apps better than others and should be standardized. 
Nonetheless, researchers have found \VT{} to be dynamic in terms of the labels given by the scanners it uses
 ~\cite{peng2019opening,miller2016reviewer,mohaisen2014av}, which affects threshold-based labeling strategies as follows. 
Threshold values that used to yield the most accurate labels might change in the future as \VT{} changes the scanners it includes in its scan reports. 
Using out-of-date or inaccurate thresholds alters the distribution of malicious and benign apps in the same dataset, effectively yielding different detection results as revealed by recent results \cite{pendlebury2019,salem2018poking}. 
On the one hand, researchers might dismiss promising detection approaches, because they underperform on a dataset that utilizes a labeling strategy that does not reflect the true nature of the apps in the dataset. 
On the other hand, developers of inadequate detection methods might get a false sense of confidence in the detection capabilities of their detection methods because they perform well, albeit using an inaccurate labeling strategy ~\cite{pendlebury2018enabling,sanders2017garbage}. 

Until a more stable alternative to \VT{} is introduced, the research community will continue to use \VT{} to label apps using subjective thresholds. 
So, the overarching objective of this paper is to provide the research community with actionable insights about \VT{}'s dynamicity, its limitations, and how to optimally interpret its scan reports to label apps accurately using threshold-based labeling strategies. 
To achieve this objective, we focus on Android apps as a case study and utilize four datasets of 53K Android malicious and benign apps. 
Furthermore, based on the identified limitations of \VT{}, we provide a blueprint for a platform that mitigates the limitations of \VT{} and provides the research community with a more stable, reliable alternative to \VT{}. 

\textbf{The contributions} of this paper, therefore, are:
\begin{itemize}
    \setlength\itemsep{0em}

	\item The dynamicity of \VT{} is common knowledge within the research community and has been mentioned in previous research without, to the best of our knowledge, being adequately discussed. In this paper, we \textbf{reveal the details of such dynamicity} and how it manifests, how it projects an improper image of the performance of otherwise competent scanners, and how it undermines the performance of threshold-based labeling strategies over time (\autoref{subsec:threshold_sensitivity}).

	\item We provide the research community with \textbf{actionable insights} about how to use threshold-based labeling strategies to label Android apps in a manner that copes with \VT{}'s dynamicity and limitations (\autoref{subsec:threshold_optimal}). We demonstrate that the optimal thresholds that yield the best labeling accuracies change over time and, hence, advise researchers to find the current optimal threshold(s) to use prior to labeling apps in datasets they use to train and evaluate malware detection methods.
    
	\item There are voices within the research community that calls for the replacement of \VT{}. However, without a clear enumeration of the shortcomings of \VT{}, we risk implementing alternative labeling platforms that suffer from the same shortcomings of \VT{}. We \textbf{identified four limitations} of \VT{} that undermine its reliability and usefulness. Those limitations are (a) frequent inclusion and exclusion of scanners in the scan reports of apps, (b) using inadequate versions of scanners that are designed to detect malicious apps for other platforms, (c) refraining from frequently and automatically reanalyzing and re-scanning apps, and (d) denying access to the history of scan reports. 
	
	\item Based on the identified limitations, we provide the community in \autoref{sec:eleda} with a blueprint of \textbf{how to build alternatives} to \VT{} that mitigate the platform's limitations. 

    
\end{itemize}
\section{Preliminaries}
\label{sec:preliminaries}
To motivate the need for our paper and its line of research, in this section, we give an example of how the dynamicity of \VT{} impacts the labeling accuracy of threshold-based labeling strategies that are commonly used within the research community.
This example also demonstrates the impact of inaccurate labeling on the reliability of malware detection methods and their results. 
Based on this example, we postulate research questions meant to reveal insights about \VT{}'s dynamicity and limitations and their impact on the labeling accuracy of threshold-based labeling strategies. 

\subsection{Motivating Example}
\label{subsec:preliminaries_example}
In this example we focus on \gls{ml}-based detection methods given their popularity within the academic community \cite{pendlebury2018enabling,suarez2017droidsieve,tam2017evolution,yang2017malware,arshad2016android}. 
Researchers devise new techniques to extract features from Android apps and use those features to train \gls{ml} models.
The trained models are evaluated by assessing their abilities to recognize the malignancy of apps not used during the training process (i.e., out-of-sample apps), by deeming them as malicious or benign. 
This process requires datasets of Android \gls{apk} archives that researchers often acquire from online repositories, such as \emph{AndroZoo} ~\cite{allix2016androzoo} or \VT{} itself. 
To simulate this process, we downloaded a random collection of 6,172 apps developed in between 2018 and 2019 from \emph{AndroZoo}. 
We refer to this dataset as \emph{AndroZoo} throughout this paper. 

The acquired \gls{apk} archives and their corresponding apps need to be labeled either as malicious and benign or in terms of the malware families and types they belong to ~\cite{wei2017deep,hurier2016lack}. 
Since manually analyzing thousands of apps is infeasible, researchers usually download the scan reports of apps in their training datasets from \VT{} and label them according to some labeling strategy they deem accurate. 
The features extracted from the downloaded apps are used alongside their labels to train \gls{ml} models. 
As an example of an \gls{ml}-based Android malware detection method, we utilize a detection method that is renowned in the research community and has been used by different researchers as a benchmark ~\cite{pendlebury2019}, namely \emph{Drebin} ~\cite{arp2014drebin}. 
In this example, we use different threshold-based labeling strategies that have been utilized by researchers in the past, viz.\ we use thresholds that deem any given app as malicious if the number of scanners in its \VT{} scan report deeming it malicious (i.e., \emph{positives}) is at least one scanner ~\cite{li2017understanding}, four scanners ~\cite{pendlebury2019,miller2016reviewer}, ten scanners ~\cite{wang2018beyond}, 50\% of scanners (i.e., $\frac{positives}{total}\geq 50\%$) ~\cite{wei2017deep}, and the strategy adopted by Arp et al.\ in ~\cite{arp2014drebin}\footnote{An app is deemed as malicious if at least two out of the following ten \VT{} scanners label it as such: \texttt{AVG}, \texttt{Avira} (formerly AntiVir), \texttt{BitDefender}, \texttt{ClamAV}, \texttt{ESET-NOD32}, \texttt{F-Secure}, \texttt{Kaspersky}, \texttt{McAfee}, \texttt{Panda}, and \texttt{Sophos}.}. 
We refer to those strategies as \texttt{vt$\geq$1}, \texttt{vt$\geq$4}, \texttt{vt$\geq$10}, \texttt{vt$\geq$50\%}, and \texttt{drebin}, respectively.

To test the ability of the trained \emph{Drebin} classifiers to classify out-of-sample apps accurately, we use two small datasets that we refer to as \emph{Hand-Labeled}\footnote{http://tiny.cc/95bhaz} and \emph{Hand-Labeled 2019}\footnote{http://tiny.cc/a7bhaz}. 
Both datasets comprise 100 Android apps that were downloaded from \emph{AndroZoo} and \emph{manually analyzed and labeled}, to acquire reliable ground truth.
The primary difference between both datasets is that apps in the latter were developed in 2019.
Furthermore, we ensured that apps in both datasets do not overlap with apps in the \emph{AndroZoo} dataset. 

\begin{figure*}
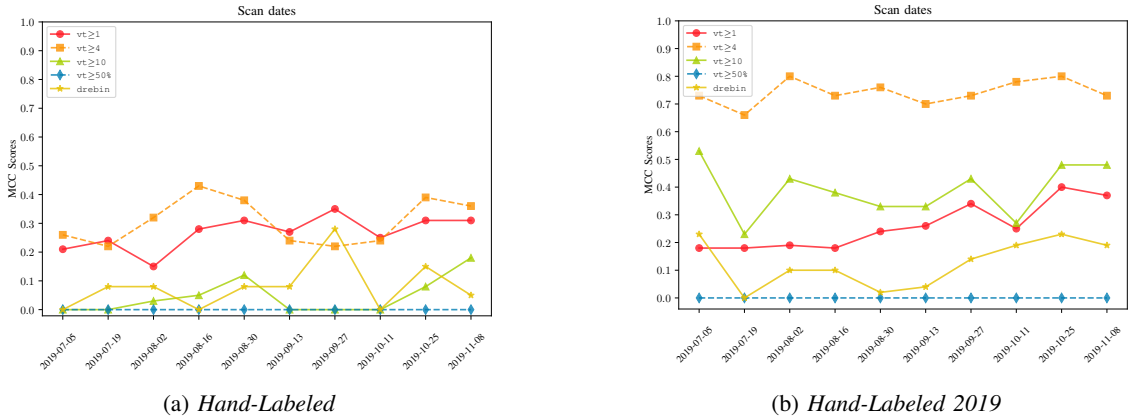

	\centering
	\begin{subfigure}{0.47\textwidth}
		\centering
		\scalebox{0.42}{\input{figures/Line_Detection_Drebin_Hand-Labeled.pgf}}
		\caption{\emph{Hand-Labeled}}
		\label{fig:detection_drebin_hand-labeled}
	\end{subfigure}
	\begin{subfigure}{0.45\textwidth}
		\centering
		\scalebox{0.42}{\input{figures/Line_Detection_Drebin_Hand-Labeled2019.pgf}}
		\caption{\emph{Hand-Labeled 2019}}
		\label{fig:detection_drebin_hand-labeled2019}
	\end{subfigure}
	\caption{The \gls{mcc} scores achieved by the \emph{Drebin} classifiers labeled using different threshold-/\gls{ml}-based labeling strategies against the \emph{Hand-Labeled} and \emph{Hand-Labeled 2019} dataset between July 5$^{th}$, 2019 and \futureDate{}.}
	\label{fig:detection_drebin}
\end{figure*}

In \autoref{fig:detection_drebin}, we plot the classification accuracy of the \emph{Drebin} classifiers whose feature vectors were labeled using different threshold-based labeling strategies over a period of four months. 
Each point on the X-axis refers to a point in time in which we re-scanned all apps in the \emph{AndroZoo}, \emph{Hand-Labeled}, and \emph{Hand-Labeled 2019} datasets on \VT{}, downloaded their up-to-date scan reports, re-trained the \emph{Drebin} classifiers using all the different labeling strategies, and re-tested the trained classifiers. 
We use the \gls{mcc} score ~\cite{sklearn2019mcc} to represent the classification accuracy of the \emph{Drebin} classifiers instead of conventional metrics, such as accuracy ~\cite{sklearn2019accuracy}, that are unable to capture or penalize bias towards certain classes in imbalanced datasets. 
The \gls{mcc} values range from -1 (i.e., all apps were misclassified) to 1 (i.e., perfect classification), with the value of 0 indicating a classification ability similar to random classification.

We are not concerned with the absolute performance of the \emph{Drebin} classifiers. 
With this example, we wish to demonstrate two issues. 
Firstly, despite all being utilized within previous research efforts, it appears that some labeling strategies (e.g., \texttt{vt$\geq$4}), contribute to training \emph{Drebin} classifiers that generalize better to out-of-sample Android apps than other strategies. 
That is using the exact same feature set and \gls{ml} algorithm, different labeling strategies significantly alters the performance of the same detection method, which leads to two different answers to the question: \emph{Is the \emph{Drebin} methods a reliable Android malware detection method?}
A researcher that opts to use the \texttt{vt$\geq$50\%} labeling strategy will deem the \emph{Drebin} method as unreliable and dismiss it, whereas one that uses the \texttt{vt$\geq$4} strategy might deem it as potentially reliable and continues to refine it. 
In general, having multiple perspectives on the reliability of a detection method might either force researchers to dismiss promising methods that underperform during the evaluation, or adopt ones that are mediocre yet perform well during the evaluation phase.
Secondly, the \gls{mcc} scores of the \emph{Drebin} classifiers appear to fluctuate from one scan date to another, although they are merely two weeks apart. 
So, depending on the scan date of the \VT{} scan reports used to label apps in the training and test datasets, researchers might get different classification accuracies from their detection methods. 
For example, during the period between September 13$^{th}$, 2019, and October 11$^{th}$, 2019, using the labeling strategy \texttt{vt$\geq$1} led to training \emph{Drebin} classifiers that performed better than other labeling strategies on the \emph{Hand-Labeled} dataset.

\subsection{Research Questions}
\label{subsec:preliminaries_rqs}
In the previous example, we learned that some threshold-based labeling strategies contribute to training \gls{ml} models that can classify out-of-sample Android apps more accurately, despite the fact that all of the strategies that we used in the example were devised by researchers and utilized within the literature. 
Furthermore, we found that the labeling accuracy of most labeling strategies seems to fluctuate over time, due to some unknown aspect of \VT{}'s dynamicity. 
This fluctuation means that threshold-based labeling strategies cannot be permanently and universally utilized to label Android apps and train \gls{ml}-based detection methods. 
In this paper, as mentioned in \autoref{sec:introduction}, we aim to provide the research community with actionable insights about the impact of \VT{}'s dynamicity on the labeling strategies they utilize to label Android apps used to train and evaluate their malware detection methods, and how this dynamicity might impact the reliability of their methods. 
We provide those insights by (1) identifying the aspects of \VT{}'s dynamicity and how it manifests, (2) finding methods to workaround such dynamicity given the lack of better alternative platforms to \VT{} and the infeasibility of manually-analyzing apps, (3) pinpointing the limitations of \VT{} that cannot be mitigated, and finally (4) using the identified limitations to sketch a blueprint for a more reliable alternative platform. 
We attempt to address these issues by answering the following research questions:
\hypertarget{text:rqs}{}
\begin{itemize}
	\itemsep0em 
	\item[\textbf{RQ1}] \RQone
	\item[\textbf{RQ2}] \RQtwo
	\item[\textbf{RQ3}] \RQthree
\end{itemize}
\section{Threshold-Based Labeling Strategies}
\label{sec:threshold}

\subsection{Accuracy of Threshold-Based Labeling Strategies}
\label{subsec:threshold_accuracy}

In this section, we discuss the ability of different threshold-based labeling strategies to accurately label apps in our test datasets, namely \emph{Hand-Labeled} and \emph{Hand-Labeled 2019}. 
Using the scan reports of apps in the aforementioned datasets downloaded at different points in time, we label the apps use different labeling strategies and compare them against the ground truth we generated by manually analyzing those apps. 
In addition to the labeling strategies we used in \autoref{subsec:preliminaries_example}, we use all thresholds between one and ten scanners and the threshold of 25\% of scanners.  

\begin{figure*}
\centering
\begin{subfigure}{0.47\textwidth}
	\centering
    \scalebox{0.42}{\input{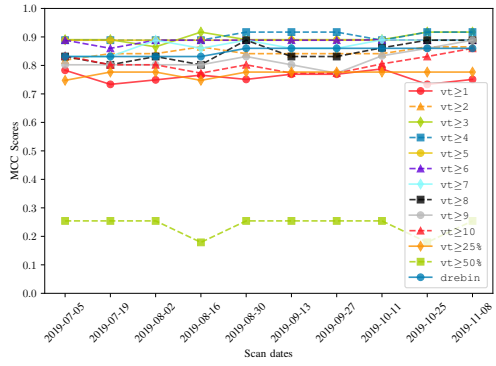}}
    \caption{\emph{Hand-Labeled}}
    \label{fig:threshold_based_accuracy_hand-labeled}
\end{subfigure}
\begin{subfigure}{0.47\textwidth}
	\centering
    \scalebox{0.42}{\input{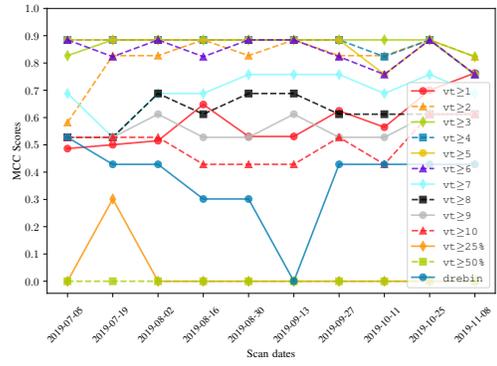}}
    \caption{\emph{Hand-Labeled 2019}}
    \label{fig:threshold_based_accuracy_hand-labeled2019}
\end{subfigure}
\caption{The labeling accuracy of different threshold-based labeling strategies against apps in \emph{Hand-Labeled} and \emph{Hand-Labeled 2019} datasets based on their \VT{} scan reports downloaded between July 5$^{th}$, 2019 and \futureDate{}. Accuracy is calculated in terms of the \gls{mcc} of each labeling strategy.}
\label{fig:threshold_based_accuracy}
\end{figure*} 

In \autoref{fig:threshold_based_accuracy}, we plot the performance of each labeling strategy on the \emph{Hand-Labeled} and \emph{Hand-Labeled 2019} datasets between July 5$^{th}$, 2019 and \futureDate{} in terms of the \gls{mcc} score. 
We attempted to gain access to the \VT{} scan reports of apps in both datasets that pre-dates July. 
Unfortunately, access to such reports is not available under academic licenses and requires the purchase of costly commercial ones, which we consider the first limitation of \VT{}. 

\hypertarget{text:lim1}{}
\begin{tcolorbox}[colback=white,title={\footnotesize \VT{} Limitation 1}]
	\footnotesize
	\VT{} does not grant access to academic researchers to the history of scan reports of apps previously added and scanned on the platform, even if such apps
were added by the academic community itself. In fact, we are not sure whether \VT{} keeps or discards the current scan reports of apps prior to rescanning apps.
\end{tcolorbox}

One can notice that some threshold-based labeling strategies are more accurate than others over time. 
Starting with the performance of \texttt{vt$\geq$1}, while the labeling strategy managed to achieve a decent \gls{mcc} score on apps in the \emph{Hand-Labeled} dataset, its performance noticeably decreased against apps in the newer \emph{Hand-Labeled 2019} dataset. 
Low threshold values, such as one or two scanners, might result in false positives, especially against new apps whose \VT{} scan reports are not mature enough. 
In most cases, if an app has one or two \VT{} scanners deeming it as malicious, it is a case of a subjective definition of malignancy. 
For example, we noticed that some scanners such as \texttt{Tencent} consistently label any apps (e.g., \href{http://tiny.cc/l945jz}{\texttt{ed23237e34ff47580a99ac70f35e84b32c05ab1d}}), that utilize \emph{App Inventor}\footnote{App Inventor is a visual programming environment maintained by MIT that enables non-technical users to develop apps for Android \cite{appinventor2019}.} as malicious apps belonging to the \texttt{A.gray.inventor.a} malware family. 

As for (\texttt{vt$\geq$50\%}), pushing the threshold that high might prevent recently-developed malicious apps and apps that belong to ambiguous malware types (e.g., \texttt{Adware}) from being labeled as malicious, resulting in a high number of false negatives. 
Similar to \texttt{vt$\geq$1}, the older the app and its \VT{} scan report, the better the performance of \texttt{vt$\geq$50\%} and the newer the app, the worse the performance, especially since the malicious apps were not deemed labeled by enough \VT{} scanners to make the 50\% mark required by the strategy to deem them as malicious successfully. 

Another aspect of how the age of apps and, in turn, the maturity of their \VT{} scan reports impacts the performance of different threshold-based labeling strategies can be seen in the proximity of different \gls{mcc} lines in \autoref{fig:threshold_based_accuracy}. 
In particular, in \autoref{fig:threshold_based_accuracy_hand-labeled}, the lines of almost all threshold-based labeling strategies are close to one another and exhibit a relatively steady performance (i.e., the performance does not noticeably fluctuate). 
However, the \gls{mcc} lines in \autoref{fig:threshold_based_accuracy_hand-labeled2019} are more distributed across the figure and exhibit more fluctuations in performance. 
For example, on the \emph{Hand-Labeled 2019} dataset, the \gls{mcc} score of \texttt{drebin} sharply decreased from a little above 0.3 on August 30$^{th}$, 2019 to almost 0.0 on September 13$^{th}$, 2019 only to sharply increase to around 0.45 two weeks later. 
The reason behind the proximity in the case of apps in the \emph{Hand-Labeled} dataset is that their \emph{positives} values are high enough to accommodate threshold values up to at least 15 scanners, which is represented by the \texttt{vt$\geq$25\%} labeling strategy. 
The novelty of malicious apps in the \emph{Hand-Labeled 2019} means that their \emph{positives} values are much lower in comparison, which prevents thresholds higher than six scanners from achieving high \gls{mcc} scores. 

\begin{figure*}
\centering
\begin{subfigure}{0.32\textwidth}
	\centering
    \scalebox{0.30}{\input{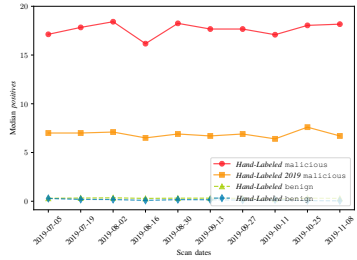}}
    \caption{Mean \emph{positives}}
    \label{fig:mean_positives}
\end{subfigure}
\begin{subfigure}{0.32\textwidth}
	\centering
    \scalebox{0.30}{\input{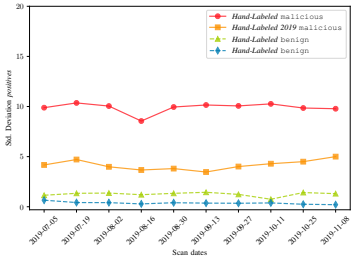}}
    \caption{Std. Deviation \emph{positives}}
    \label{fig:mean_positives}
\end{subfigure}
\begin{subfigure}{0.30\textwidth}
	\centering
    \scalebox{0.30}{\input{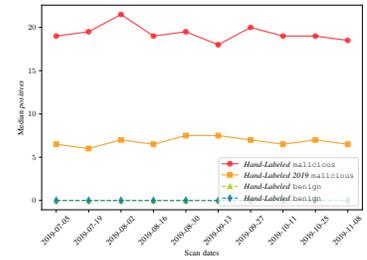}}
    \caption{Median \emph{positives}}
    \label{fig:median_positives}
\end{subfigure}
\caption{The mean, standard deviation, and median of the \emph{positives} attributed found in scan reports of apps in the \emph{Hand-Labeled} and \emph{Hand-Labeled 2019} datasets between July 5$^{th}$, 2019 and \futureDate{}.}
\label{fig:mean_median_std_positives}
\end{figure*} 

To corroborate this argument, we calculated the mean, median, and standard deviation of the \emph{positives} attribute for apps in both datasets over the same period of time. 
The results in \autoref{fig:mean_median_std_positives} show that the \emph{positives} attribute of malicious apps in the \emph{Hand-Labeled} dataset stays within the range of 15 to 20. 
Even with a standard deviation of ten scanners, the range of scanners needed to label malicious apps in this dataset correctly remains between 7 and 20 scanners. 
As for the malicious apps in the \emph{Hand-Labeled 2019} dataset, their \emph{positives} values have mean and median values around seven, ranging between 2.89 scanners and 10.91 scanners. 
The benign apps in both datasets have mean and median values that are almost zero with a negligible standard deviation of at most one scanner. 
So, any threshold values above three are guaranteed to avoid false positives resulting from deeming benign apps as malicious. 
The aforementioned values of the \emph{positives} attribute allows the labeling strategies using thresholds between three and six \VT{} scanners (i.e., \texttt{vt$\geq$3}, \texttt{vt$\geq$4}, \texttt{vt$\geq$5}, and \texttt{vt$\geq$6}), to outperform all other threshold-based labeling strategies on both datasets in terms of the \gls{mcc}. 

What we can conclude from this measurement is that the thresholds that result in decent labeling accuracy differ from one dataset to another depending on one the age of apps in the dataset and the maturity of their \VT{} scan reports. 
To generalize those thresholds to multiple datasets, one must possess a diverse dataset that includes Android apps of different ages, which is a concept we discuss in detail in \autoref{subsec:threshold_optimal}.  

\subsection{Sensitivity to \VT{}'s Dynamicity}
\label{subsec:threshold_sensitivity}
In the previous section, despite finding that a range of thresholds between three and six yields the best \gls{mcc} scores on the \emph{Hand-Labeled} and \emph{Hand-Labeled 2019} datasets, we noticed that the performance of labeling strategies utilizing these thresholds fluctuates at different points in time especially against the latter dataset. 
This fluctuation is largely attributed to \VT{}'s dynamicity and the immaturity of the scan reports of recently-developed Android apps. 
In this section, we analyze (a) whether the reason behind such fluctuation is indeed the dynamicity of \VT{}, and (b) the aspects of such dynamicity that cause this fluctuation. 
In this analysis, we focus on the performance of threshold-based labeling strategies on the \emph{Hand-Labeled 2019} dataset on two dates, namely \latestDate{} and \futureDate{}. 

We found that the labeling accuracy of threshold-based labeling strategies using thresholds between three and six on benign apps in the \emph{Hand-Labeled 2019} dataset did not change, according to their specificity scores ($\frac{TN}{N}$). 
Focusing on their performance on the malicious apps, we found that they respectively had recall scores ($\frac{TP}{P}$) of 0.7, 0.6, and 0.6, 0.6 on \futureDate{} instead of 0.8, 0.8, 0.8, and 0.7 on \latestDate{}. 
Since the total number of malicious apps in this dataset is ten, we can investigate the differences in \emph{positives} values in their scan reports and the different \VT{} scanners that deemed them malicious on both dates. 
In \autoref{tab:handlabeled_2019_malicious_evolution}, we detail the change in the \emph{positives} values in terms of the \VT{} scanners that deemed the apps as malicious which (a) were added to the scan reports on \futureDate{}, (b) were removed from the \latestDate{} scan reports, and (c) changed their verdicts between both dates.  

\bgroup
\def\arraystretch{1.0}
\begin{table*}[]
\centering
\caption{The evolution of \emph{positives} for apps in the \emph{Hand-Labeled 2019} dataset that we deemed malicious after manual analysis and a detailed view of the \VT{} scanners that were added/removed or changed their verdicts between \latestDate{} and \futureDate{} and how that affected the performance of \texttt{vt$\geq$3}, \texttt{vt$\geq$4}, \texttt{vt$\geq$5}, and \texttt{vt$\geq$6}. The check mark (\checkmark) depicts whether the threshold-based labeling strategy managed to detect the malicious app on \futureDate{}.}
\label{tab:handlabeled_2019_malicious_evolution}
\tiny
\resizebox{\textwidth}{!}{
\begin{tabular}{@{}|c|cccccccccc|@{}}
\toprule
App's \texttt{SHA1} Hash & \begin{tabular}[c]{@{}c@{}}\emph{positives}\\ (\latestDate{})\end{tabular} & \begin{tabular}[c]{@{}c@{}}\emph{positives}\\ (\futureDate{})\end{tabular} & \begin{tabular}[c]{@{}c@{}}Added\\ Positives\end{tabular} & \begin{tabular}[c]{@{}c@{}}Removed\\ Positives\end{tabular} & \begin{tabular}[c]{@{}c@{}}Flipped to\\ Positive\end{tabular} & \begin{tabular}[c]{@{}c@{}}Flipped to\\ Negative\end{tabular} & \rotatebox[origin=c]{90}{ \texttt{vt$\geq$3} } & \rotatebox[origin=c]{90}{ \texttt{vt$\geq$4} } & \rotatebox[origin=c]{90}{ \texttt{vt$\geq$5} } & \rotatebox[origin=c]{90}{ \texttt{vt$\geq$6} } \\ \hline
\href{http://tiny.cc/j6m5jz}{\texttt{8a9...}} & 0 & 0 & -- & -- & -- & -- &  &  &  & \\ \hline
\href{http://tiny.cc/2an5jz}{\texttt{bd9...}} & 5 & 3 & -- & \begin{tabular}[c]{@{}c@{}}\texttt{ESET-NOD32}\\ \texttt{Fortinet}\\ \texttt{Ikarus}\end{tabular} & \texttt{Cyren} & -- & \checkmark &  &  & \\ \hline
\href{http://tiny.cc/hen5jz}{\texttt{765...}} & 7 & 7 & -- & -- & \texttt{Zillya} & \texttt{Trustlook} & \checkmark & \checkmark & \checkmark & \checkmark\\ \hline
\href{http://tiny.cc/ufn5jz}{\texttt{5be...}} & 12 & 13 & \texttt{Ikarus} & -- & -- & -- & \checkmark & \checkmark & \checkmark & \checkmark \\ \hline
\href{http://tiny.cc/jgn5jz}{\texttt{6da...}} & 7 & 6 & -- & -- & -- & \texttt{Trustlook} & \checkmark & \checkmark & \checkmark & \checkmark \\ \hline
\href{http://tiny.cc/4kn5jz}{\texttt{c70...}} & 13 & 13 & -- & -- & \texttt{Symantec} & \texttt{Zoner} & \checkmark & \checkmark & \checkmark & \checkmark \\ \hline
\href{http://tiny.cc/ghn5jz}{\texttt{b9f...}} & 10 & 12 & \texttt{Ikarus} & -- & \begin{tabular}[c]{@{}c@{}}\texttt{AegisLab}\\ \texttt{K7GW}\end{tabular} & \texttt{McAfee} & \checkmark & \checkmark & \checkmark & \checkmark \\ \hline
\href{http://tiny.cc/dnn5jz}{\texttt{a0a...}} & 8 & 11 & \texttt{Fortinet} & -- & \begin{tabular}[c]{@{}c@{}}\texttt{Zillya}\\ \texttt{Zoner}\end{tabular} & -- & \checkmark & \checkmark & \checkmark & \checkmark \\ \hline
\href{http://tiny.cc/gon5jz}{\texttt{90e...}} & 6 & 1 & -- & \begin{tabular}[c]{@{}c@{}}\texttt{ESET-NOD32}\\ \texttt{Fortinet}\\ \texttt{Ikarus}\\ \texttt{Yandex}\end{tabular} & -- & \texttt{Cyren} &  &  &  & \\ \hline
\href{http://tiny.cc/fpn5jz}{\texttt{0d5...}} & 1 & 1 & -- & -- & -- & -- &  &  &  & \\ \hline
\multicolumn{7}{|c|}{Recall} & 0.70 & 0.60 & 0.60 & 0.60 \\
\multicolumn{7}{|c|}{\gls{mcc}} & 0.82 & $\approx$0.76 & $\approx$0.76 & $\approx$ 0.76 \\ \bottomrule
\end{tabular}}
\end{table*}
\egroup

Three apps out of ten did not encounter any change in their \emph{positives} values. 
Nonetheless, one\footnote{\href{http://tiny.cc/lzm5jz}{7658f70ae6acccfa9f3e900f8ae689603cc19d0b}} of these apps maintained the same value of \emph{positives} because two \VT{} scanners changed their verdicts, namely \texttt{Zillya} changed its verdict from benign to malicious and \texttt{Trustlook} changed its verdict vice versa. 
However, \VT{} contributed to altering the performance of all labeling strategies by removing scanners that correctly deemed two apps malicious from their scan reports. 
In particular, the scanners \texttt{ESET-NOD32}, \texttt{Fortinet}, and \texttt{Ikarus} were removed from one\footnote{\href{http://tiny.cc/a2m5jz}{bd97c85d38bd5bfc5e29b05b1a3a81b12949065a}} app's scan report, effectively reducing its \emph{positives} value from five on \latestDate{} to only three on \futureDate{}; this prevented the \texttt{vt$\geq$4}, \texttt{vt$\geq$5}, and \texttt{vt$\geq$6} labeling strategies from correctly deeming the app as malicious. 
The same scanners along with \texttt{Yandex} were not included in the second\footnote{\href{http://tiny.cc/g3m5jz}{90e6ac481fdd497f152234f1cd5bec6d40f50037}} app's \futureDate{} scan report, which brought the value of \emph{positives} from six to one, putting the app beyond the reach of all of the three aforementioned strategies. 

Unfortunately, we cannot identify the reasons behind \VT{}'s decision to alter the set of scanners it includes in an app's scan report. 
It is a puzzling fact that \VT{} added the \texttt{Ikarus} and \texttt{Fortinet} scanners to the scan reports of some apps, whilst removing them from the reports of others. 
The more confusing fact is that the removed scanners contain the \texttt{ESET-NOD32} scanner, which we found to have correctly labeled the apps as malicious on \latestDate{}. 
This seemingly haphazard inclusion and exclusion of scanners within periods as small as two months contribute to the frequent change of the currently optimal range of thresholds. 
While such change does not impact older apps, such as the ones in the \emph{Hand-Labeled} dataset, its impact is noticeable on newer apps, such as the ones in the \emph{Hand-Labeled 2019} dataset. 
With this set of measurements, we unveil the first limitation of \VT{}, which has a direct impact on the performance of threshold-based labeling strategies, viz.:

\hypertarget{text:lim2}{}
\begin{tcolorbox}[colback=white,title={\footnotesize \VT{} Limitation 2}]
	\footnotesize
    \VT{} changes the set of scanners it includes in the scan reports of apps over time by including and excluding the verdicts of scanners regardless of the quality of those verdicts.
\end{tcolorbox}

During this analysis, we noticed that the \VT{} version of some of the renowned scanners, such as \texttt{BitDefender} and \texttt{Panda}, fail to recognize the malignancy of any of the malicious apps in either dataset. 
Both of those scanners continue to be given good reviews by users on the Google Play marketplace and, more importantly, on platforms that assess the effectiveness of antiviral software such as \emph{AV-Test} ~\cite{avtest2019}. 
\VT{} states that the versions of scanners it uses "\emph{may differ from commercial off-the-shelf products. The [antiviral software] company decides the particular settings with which the engine should run in VirusTotal}" ~\cite{virustotal2019}. 
In fact, we found that the version used by \VT{}'s for \texttt{BitDefender}, for instance, is 7.2, whereas the versions available on Google Play have codes between 3.3 and 3.6. 
The 7.2 version of \texttt{BitDefender} corresponds to a free edition version developed for Windows-based malware that targets older versions of Windows, such as \texttt{Windows XP} ~\cite{pcmagazin2008bitdefender} and, hence, is inadequate to use to detect Android malware. 
The positive reputation that \texttt{BitDefender} has in the market suggests that using its adequate version (i.e., the one that is designed to detect Android malware), would yield a detection performance better than the version on \VT{}. 
To verify this hypothesis, we downloaded and installed the latest version of the \texttt{BitDefender} scanner from the Google Play marketplace, installed it on an \gls{avd}, and used it to scan the malicious apps in both the \emph{Hand-Labeled} and \emph{Hand-Labeled 2019} datasets. 
We also downloaded ten apps randomly sampled from the \emph{AMD} ~\cite{wei2017deep} dataset to test \texttt{BitDefender}'s accuracy. 
Unlike the results obtained from \VT{} that the scanner detected \textbf{none} of those malicious apps, we found that \texttt{BitDefender} detects 56.5\% of the malicious apps in the \emph{Hand-Labeled} dataset, 20\% of those in the \emph{Hand-Labeled 2019} dataset, and 70\% of those in the sampled \emph{AMD} dataset. 
Figuring out the reason why antiviral software companies opt to provide \VT{} with older, inadequate versions of their scanners is not in the scope of this paper and, in fact, impossible to answer on behalf of antiviral software companies.  
However, it leads us to identify the third limitation of \VT{}: 

\hypertarget{text:lim3}{}
\begin{tcolorbox}[colback=white,title={\footnotesize \VT{} Limitation 3}]
	\footnotesize
    \VT{} may replace the versions of scanners with inadequate ones that are not designed to detect Android malware presumably based on the request of the scanner's vendor or managing firm.
\end{tcolorbox}

\subsection{Finding the Optimal Threshold}
\label{subsec:threshold_optimal}
In the previous section, we found that \VT{} changes the set and versions of scanners it includes in the scan reports of apps.
This change impacts the long term labeling accuracy of threshold-based labeling strategies. 
For example, while the dynamicity of \VT{} has caused the \gls{mcc} score of \texttt{vt$\geq$4} to decrease against the \emph{Hand-Labeled 2019} dataset from 0.89 to 0.76 (i.e., a decrease of 14.61\%), it did not have an impact on the \gls{mcc} scores of \texttt{vt$\geq$2} yet caused the scores of \texttt{vt$\geq$1} to decrease. 
Effectively, the previously-discussed aspects of \VT{}'s dynamicity cause threshold-based labeling strategies to trade places in terms of the most accurate ones.  
In other words, at any given moment in time, a (different) subset of threshold-based labeling strategies will depict the most accurate labeling strategies (i.e., optimal thresholds). 
So, before labeling apps in their training and evaluation datasets, researchers must identify the most accurate threshold(s) at that particular point in time.
Given a reference set of Android apps whose ground truth is known, one straightforward method to identify the currently accurate thresholds is to download the apps' latest \VT{} scan reports, compare the labeling accuracy of all thresholds between one and 60 (i.e., average total number of scanners), and choose the thresholds that yield the best scores. 
In this section, we investigate the feasibility of this brute force approach to identify the optimal thresholds of \VT{} scanners at any point in time. 

\begin{algorithm}
\caption{An algorithm to find the current optimal threshold of \VT{} scanners to use in labeling Android apps.}
\small
\label{alg:bruteforce_threshold}
\begin{algorithmic}[1]
\Procedure{FindCurrentOptimalThreshold}{$A, \gamma$}
	\State \emph{tmpResults} = $\lbrace \rbrace$ 
	\ForAll {$\alpha\in A$} 
		\State \emph{response} = \VT{}.\textcolor{ACMBlue}{\texttt{rescanApp}}($\alpha$)
		\If {\emph{response} == \texttt{True}}
			\State \emph{report} = \VT{}.\textcolor{ACMBlue}{\texttt{downloadReport}}($\alpha$)
			\If {\emph{report} != \texttt{Null}}
				\State \emph{positives}$_\alpha$ = \emph{report}[\textcolor{ACMPurple}{"positives"}]
				\ForAll {$\sigma \in \lbrace 1,2,3,...,60\rbrace$}
					\If {\emph{positives}$_{\alpha}\geq\sigma$}
						\State \emph{label}$_{\alpha}$ = \textcolor{ACMRed}{\texttt{malicious}}
					\Else
						\State \emph{label}$_{\alpha}$ = \textcolor{ACMGreen}{\texttt{benign}}
					\EndIf
					\State \emph{tmpResults}[\textcolor{ACMPurple}{\texttt{vt$\geq\sigma$}}].\textcolor{ACMBlue}{\texttt{append}}(\emph{label}$_{\alpha}$)
				\EndFor											
			\EndIf
		\EndIf						
	\EndFor \\
	\State \emph{bestThreshold} = ""
	\State \emph{bestScore} = 0.0
	\ForAll {$\sigma \in \lbrace 1,2,3,...,60\rbrace$}
		\State \emph{currentScore} = \textcolor{ACMBlue}{\texttt{calculateScore}}(\emph{tmpResults}[\textcolor{ACMPurple}{\texttt{vt$\geq\sigma$}}], $\gamma$)
		\If { \emph{currentScore} $\geq$ \emph{bestScore} }
			\State \emph{bestScore} = \emph{currentScore}
			\State \emph{bestThreshold} = \textcolor{ACMPurple}{\texttt{vt$\geq\sigma$}}
		\EndIf
	\EndFor
		
	\Return \emph{bestThreshold}, \emph{bestScore}
\EndProcedure
\end{algorithmic}
\end{algorithm}

\autoref{alg:bruteforce_threshold} depicts a simple algorithm to find the current optimal threshold of \VT{} scanners that yields the most accurate labels. 
To assess the quality of labels given by different threshold-based labeling strategies, this algorithm requires the presence of a dataset ($A$) of pre-labeled Android malicious and benign apps. 
As mentioned earlier in this paper, the most reliable ground truth ($\gamma$) can be generated using manual analysis and labeling of apps. 
Without such a reliable ground truth ($\gamma$) that acts as a reference to compare against, one has to choose a subjective threshold ($\sigma$) that one believes represents the nature of apps in ($A$) and use it as ground truth ($\gamma_\sigma$) 
If so, the only threshold that would generate labels mimicking ($\gamma_\sigma$) would be ($\sigma$) itself. 
Effectively, we end up with the same problem we are attempting to avoid, namely that of choosing subjective thresholds based on personal views.

Relying on manual analysis already introduces infeasibility to the algorithm. 
However, assuming the existence of pre-labeled apps, another problem arises. 
As discussed earlier, the immaturity of newly-developed Android malware and the dynamicity of \VT{} lowers the values of the \emph{positives} attribute in the scan reports of those apps which, in turn, lowers the thresholds needed to label them accurately. 
Without access to newly-developed Android malware, researchers risk choosing thresholds based on the scan reports of old malicious apps, which are usually higher than the thresholds required to detect new malware. 
For example, in \autoref{fig:threshold_based_accuracy}, if a researcher only has access to the \emph{Hand-Labeled} dataset, on October 11$^{th}$, 2019, they might opt to use the \texttt{drebin} labeling strategy because it exhibits stable performance of high \gls{mcc} scores. 
However, this labeling strategy will perform much worse on newer apps in the \emph{Hand-Labeled 2019} dataset (i.e., it does not generalize to newer malicious apps). 
Consequently, researchers adopting this brute force approach to finding the currently optimal thresholds need to continuously update their reference datasets with newly-developed and discovered Android (malicious) apps. 

If the reference dataset ($A$) satisfies the previous condition, identifying the currently optimal threshold can be performed as follows. 
For each app in the dataset ($\alpha \in A$), the latest scan report of ($\alpha$) needs to be acquired. 
Firstly, the user has to issue a rescan request to \VT{}, which takes around four minutes to complete (line \texttt{5})
As discussed earlier, this request can be issued using the platform's web interface or using the \gls{api} interface. 
In general, under the academic license, a total of 20K requests can be issued per day. 
So, depending on the size of the reference dataset ($A$), the process of rescanning all apps might take several days or even months. 
Furthermore, we recently were forbidden from issuing this type of request using our academic license. 
As of the date of writing this paper, we are unaware of whether \VT{} prevents academic licenses from issuing this type of requests, or whether we are encountering an individual technical difficulty. 
We consider the decision of \VT{} not to automatically and regularly rescan apps as another limitation of the platform:

\hypertarget{text:lim4}{}
\begin{tcolorbox}[colback=white,title={\footnotesize \VT{} Limitation 4}]
	\footnotesize
	\VT{} does not rescan the apps it possesses on a regular basis and delegates this task to manual requests issued by its users. One direct consequence of this decision is prolonging the process of acquiring up-to-date scan reports of apps. 
\end{tcolorbox}

In line \texttt{7}, after the rescan requests are completed, researchers need to download the up-to-date scan reports from \VT{}. 
Similar to the rescan \gls{api} requests, download requests are limited to 20K requests per day, which might add a few more days to the process. 
Between lines \texttt{8} and \texttt{23}, the process becomes straightforward. 
Using thresholds ($\sigma$) between one and 60, the labels of apps in ($A$) are calculated and stored in a temporary structure under the key \texttt{vt$\geq\sigma$} (line \texttt{15}). 
The stored labels are then compared against the ground truth ($\gamma$), and a score is calculated, say \gls{mcc}. 
The threshold-based based strategy that yields the best score is returned to the user. 

\section{An Alternative to \VT{}}
\label{sec:eleda}
\begin{figure}
\centering
\caption{An overview of the modules and operations of the proposed \VT{} replacement, \Eleda{}.}
\label{fig:eleda_overview}
\includegraphics[scale=0.47]{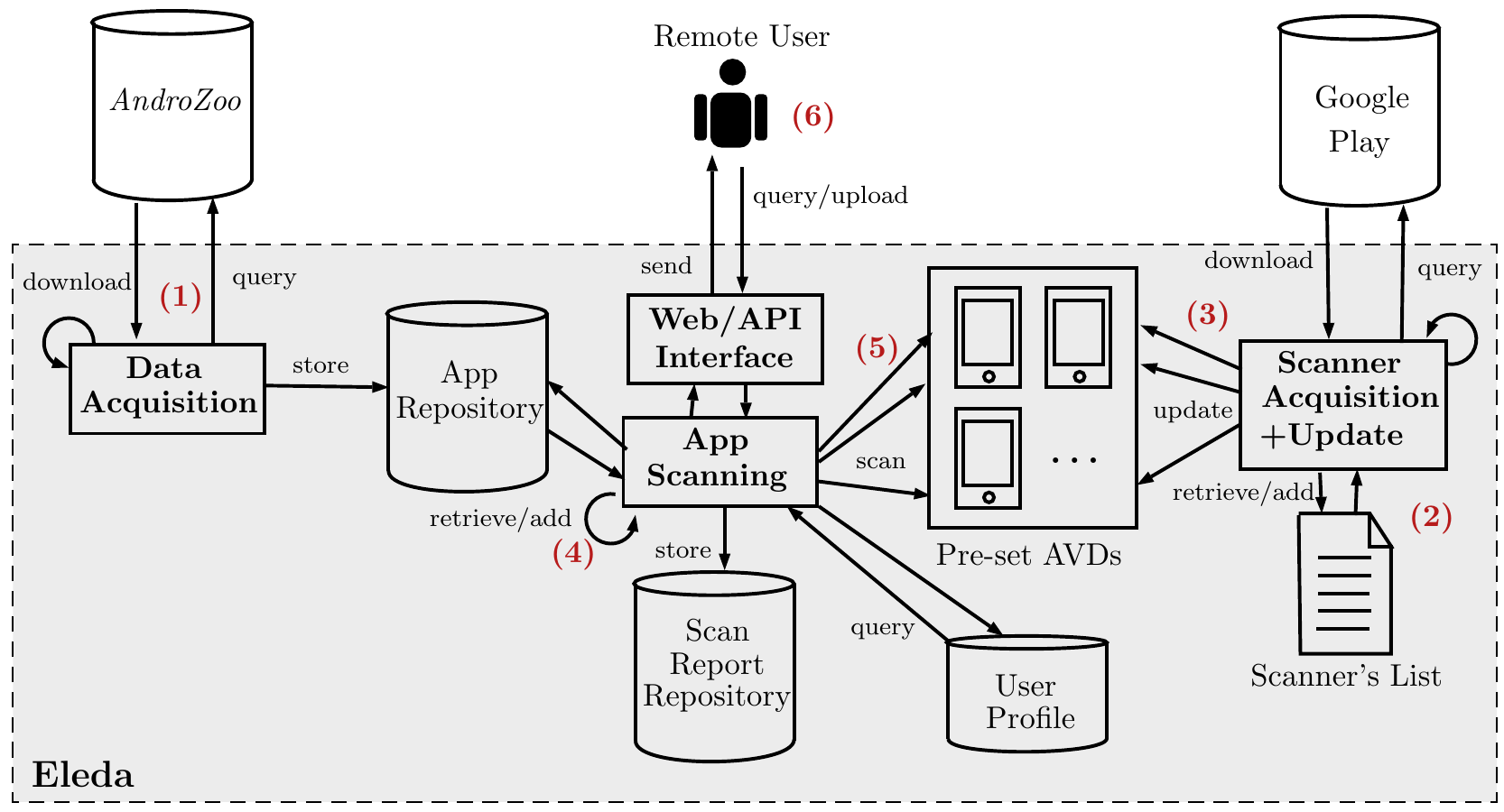}
\end{figure}

In this section, we discuss how the limitations we discussed above can be addressed upon building alternative platforms. 
We do this by describing the architecture of a hypothetical new platform, called \Eleda{}\footnote{Eleda is one of the mirror twins in Sharon Shinn's novel \emph{The Truth-Teller's Tale}, who is a truth-teller incapable of telling lies, earning her the society's trustworthiness. With Eleda's proposed design, we aspire to provide the research community with an alternative to \VT{} that is more stable and reliable.}, that is designed to mitigate \VT{}'s limitations. 
An overview of \Eleda{}'s modules and operations can be seen in \autoref{fig:eleda_overview}. 

\textbf{Data Acquisition.} 
The first operation of \Eleda{} is to acquire \gls{apk} archives of Android malicious and benign apps to scan and analyze (step \textbf{(1)}). 
\emph{AndroZoo} ~\cite{allix2016androzoo} automates the process of crawling different app marketplaces and continuously downloads their \gls{apk} archives. 
Access to such a platform is granted to researchers via an \gls{api} key that can be used to download apps using \texttt{cURL}. 
Using this module, \Eleda{} can frequently query \emph{AndroZoo} for newly-crawled and downloaded apps, which is indicated using the looped arrow. 

\textbf{Scanner Acquisition and Update.}
This module is responsible for analyzing and scanning the acquired apps. 
In step \textbf{(2)}, the module retrieves the names of antiviral scanners and queries app marketplaces, such as Google Play, for their latest versions. 
This list can be manually populated at first to include the list of \VT{} scanners that are designed to detect Android malware and are available on Google Play. 
As of March 2020, 38 ($\approx$63\%) out of around 60 scanners on \VT{} are available on Google Play. 
After downloading the latest version of each scanner in the list, \Eleda{} updates an \gls{avd} that is used to scan \gls{apk} archives (step \textbf{(3)}). 
The process of acquiring and updating new versions of antiviral scanners from Google Play--and possibly third-party Android app marketplaces--is meant to mitigate the \hyperlink{text:lim3}{third limitation} of \VT{}, namely using inadequate scanners and scanner versions not designed to scan Android apps. 
Moreover, using a pre-populated list of scanners is meant to keep the set of antiviral scanners used to scan \gls{apk} archives constant (i.e., mitigating the \hyperlink{text:lim2}{second limitation}). 

\textbf{App Scanning.}
To address the \hyperlink{text:lim4}{fourth} limitation of \VT{} that it only re-scans Android apps upon request, in step \textbf{(4)}, \Eleda{}'s App Scanning module retrieves the set of \gls{apk} archives available in the platform's app repository, scans them using the \gls{avd}s set up by the Scanner Acquisition and Update module, builds the latest scan reports of those apps, and stores them in another repository (e.g., in \gls{json} format). 
In order to make the transition from \VT{} to \Eleda{} seamless, the scan reports can contain the same information contained in \VT{} scan reports. 
For example, static information about the app components can also be extracted using analysis tools, such as \emph{Androguard} ~\cite{androguard2018}, and the \gls{api} calls issued by the app during runtime can also be monitored and recorded ~\cite{salem2019towards}. 
The frequency of the re-scan operation can be set by the users of \Eleda{}. 
Given that the platform is expected to store a large number of \gls{apk} archives, it need not re-scan all apps at the same time. 
Instead, each app can be scanned every constant interval (e.g., two weeks), starting from its initial acquisition date. 

\textbf{User Interaction.}
The last operation of \Eleda{} is user interaction. 
In step \textbf{(6)}, the platform receives a query from a remote user in the format of a hash of an Android app's \gls{apk} or the archive itself. 
If the app is not already in the platform's app repository, the App Scanning module can add the app's \gls{apk} archive to the repository. 
The \gls{apk} archive is then scanned using the platform's \gls{avd}s, and its scan report can be displayed to the user. 
\Eleda{} can mimic the design of \VT{} by offering users to interact with the platform using a web-based interface or using \gls{api}-requests. 
However, to address \VT{}'s \hyperlink{text:lim1}{first} limitation, \Eleda{} can provide the users with all scan reports of the queried app.

\section{Discussion}
\label{sec:discussion}
In this section, we discuss the insights we gained from our experiments, the limitations of our work, and how we (plan to) address them. 

\textbf{Aspects of \VT{}'s dynamicity.} In answering \hyperlink{text:rqs}{\textbf{RQ1}}, we attempted to (a) reveal the aspects of \VT{}'s dynamicity that are not clearly documented within the literature, and (b) study the impact of such dynamicity on the labeling performance of threshold-based labeling strategies. In \autoref{subsec:threshold_accuracy}, we found that \VT{} regularly manipulates the set of scanners it includes in the scan reports of apps by adding/removing scanners from such reports, including ones that correctly label the apps. This leads to changing the number of scanners that deem an app as malicious (i.e., \emph{positives}), which is what threshold-based labeling strategies hinge on to discern an app's malignancy. While this change has a negligible impact on the labels of benign apps, we found that malicious apps--especially those recently developed--suffer the most from this frequent manipulation of scanners. So, to answer \hyperlink{text:rqs}{\textbf{RQ1}}, the regular manipulation of scan reports means that thresholds that proved to be accurate at one point in time cannot be used over an extended period. 

\textbf{Optimally using \VT{} to label apps.} Given the aforementioned dynamicity of \VT{}, the concern of \hyperlink{text:rqs}{\textbf{RQ2}} is to find a workaround for the platform's dynamicity to be able to label (Android) apps based on their scan reports. We found in \autoref{subsec:threshold_accuracy} that, at any point in time, there are thresholds that provide the most accurate threshold-based labeling strategies. So, instead of relying on a fixed threshold to label apps (e.g., four scanners), for prolonged periods, researchers should identify the \emph{current optimal} thresholds based on the latest scan reports of the apps they wish to label (i.e., after re-scanning the apps on \VT{}). In \autoref{subsec:threshold_optimal}, we propose an algorithm to identify such thresholds with the help of a diverse, pre-labeled dataset of apps, and detail the possible challenges that might face researchers adopting this algorithm. 

\textbf{\VT{}'s Limitations.} The research community has long aspired to replace \VT{} with a more reliable alternative. However, without pinpointing the shortcomings of \VT{}, we risk constructing a platform that suffers from the same limitations. So, it is imperative to identify those limitations, which is the concern of \hyperlink{text:rqs}{\textbf{RQ3}}. In this paper, we identified four limitations of \VT{} that jeopardizes its usefulness. First, the platform does not grant access to the history of scans, effectively preventing researchers from studying the performance of scanners over extended periods of time. Second, the platform changes the set of scanners it uses to scan the same apps over time, which undermines the sustainability of threshold-based labeling strategies. Third, the platform uses scanners or versions of scanners that are not suitable to detect Android malware. Fourth, the platform does not automatically re-scan apps and relies on manually re-scanning apps either via its web-interface or via remote \gls{api} requests. In \autoref{sec:eleda}, we give an example of how such limitations can be mitigated upon constructing an alternative platform. 

\subsection{Limitations and Threats to Validity}
\label{subsec:threats}
\textbf{Internal validity} is concerned with actions or factors that could influence our results. In this paper, we relied on ground truth for the \emph{Hand-Labeled} and \emph{Hand-Labeled 2019} datasets that were generated after manually analyzing their apps. The intrinsic subjectivity of manually deeming apps as malicious can threaten the reliability of this ground truth. In the process of manually labeling apps in those two datasets, we complemented our process of static and dynamic analysis of apps with consulting the \VT{} scan reports of those apps. We opted to ignore the verdicts of \VT{}'s scanners vis-\`{a}-vis the verdicts of a few apps that we clearly observed their malicious behavior despite being labeled as benign by all scanners. In fact, it is not uncommon for antiviral scanners to be oblivious to the malignancy of malicious apps. For example, over the past three years, we tracked the verdicts given by \VT{} scanners to a repackaged, malicious version\footnote{\href{http://tiny.cc/ryn5jz}{aa0d0f82c0a84b8dfc4ecda89a83f171cf675a9a}} of the \texttt{K9 Mail} open source app \cite{k9mail2019} that has been developed by one of our students during a practical course. Despite being a malicious app of type \texttt{Ransom}, the scanners continued to unanimously deem the app as benign since February 8$^{th}$, 2017, even after analyzing and re-scanning the app. Another example is an app\footnote{\href{http://tiny.cc/vzn5jz}{66c16d79db25dc9d602617dae0485fa5ae6e54b2}: A calculator app grafted with a logic-based trigger that deletes user contacts only if the result of the performed arithmetic operation is 50.} that we repackaged three years ago; the app continued to be labeled as benign by all scanners until only \texttt{K7GW} recognized the app's malignancy in July 2019 and labeled it as a \texttt{Trojan}.

\textbf{External validity} focuses on the possibility of generalizing our results. There are two main aspects of generalization in our case. Firstly, our results are confined to two small datasets of Android apps. We chose to limit the size of our datasets to ensure rigorous analysis of their codebases and runtime behaviors in order to generate as reliable ground truth as possible. To compensate for their small size, we randomly downloaded 100 apps from \emph{AndroZoo} that should act as a random sample of Android apps without bias towards malignancy, category, or marketplace. The second aspect of generalization is the confinement to Android, which we adopted as a case study in this paper. So, are our results transferrable to other domains (e.g., Windows)? Concerning the \hyperlink{text:lim1}{first} and \hyperlink{text:lim4}{fourth} limitations of \VT{}, we found that the platform implements the same policies regardless of the domain researchers wish to focus on. As for the \hyperlink{text:lim2}{second} and \hyperlink{text:lim3}{third} limitations, using ten randomly sampled Windows-based malicious apps that we acquired from \VT{} (e.g., \footnote{\href{http://tiny.cc/6njwlz}{0b3beb60b80bba63154ab7491046528be0054e10}}), we found that \VT{} uses the same versions of scanners for apps belonging to different domains. We also found that \VT{} changes the set of scanners it includes in the apps' scan reports at two different scan dates. So, it seems that our insights might also generalize to domains beyond Android. 

\textbf{Reliability validity} is concerned with the possibility of replicating our findings. To replicate our results using the same datasets, we \href{http://tiny.cc/bgmapz}{offer} the scan reports of apps in the \emph{Hand-Labeled} and \emph{Hand-Labeled 2019} datasets, the \emph{Drebin} feature vectors used in \autoref{subsec:preliminaries_example}, and the tools we used to carry out our measurements and experiments to the research community.

\section{Related Work}
\label{sec:related}
We can categorize the insights and results of this paper into two categories. 
Firstly, we study \VT{} and its dynamicity, detail how this dynamicity manifests itself, and demonstrate how it impacts threshold-based labeling strategies commonly used within the research community. 
Secondly, based on our findings from studying \VT{}, we attempt provide the research community with actionable insights about how to optimally utilize \VT{} until a valid alternative is implemented. 
In this context, we surveyed the literature to find related work that fall under the categories of (a) studying \VT{}, and (b) using it for accurate labeling. 

\textbf{Studying \VT{}.}
The research community has studied different aspects of \VT{} and its scanners. 
In ~\cite{mohaisen2013towards}, Mohaisen et al.\ inspected the relative performance of \VT{} scanners on a small sample of manually-inspected and labeled Windows executables. 
The authors introduced four criteria, called correctness, completeness, coverage, and consistency, to assess the labeling capabilities of \VT{} scanners and demonstrated the danger of relying on \VT{} scanners that do not meet such criteria. 
The main objective of this study is, therefore, to shed light on the inconsistencies among \VT{} scanners on a small dataset.
In ~\cite{mohaisen2014av}, Mohaisen and Alrawi built on their previous study and attempted to assess the detection rate, the correctness of reported labels, and the consistency of detection of \VT{} scanners according to the aforementioned four criteria. 
They showed that in order to obtain complete and correct (i.e., in comparison to ground truth) labels from \VT{}, one needs to utilize multiple independent scanners instead of hinging on one or a few of them.
Similarly, within the domain of Android malware, Hurier et al.\ studied the scan reports of \VT{} scanners to identify the lack of consistency in labels assigned to the same app by different scanners and proposed metrics to quantitatively describe such inconsistencies ~\cite{hurier2016lack}. 
More recently, Peng et al.\ ~\cite{peng2019opening} showed that \VT{} scanners exhibit similar inconsistencies upon deeming \gls{url} as malicious and benign. 
The authors also showed that some \VT{} scanners are more correct than others, which requires a strategy to label such \gls{url}s that does not treat all scanners equally. 
While these studies revealed inconsistencies in the verdicts given by \VT{}, they did not delve into the possible reasons behind such inconsistences (i.e., whether they are indeed due to the scanners' incompetences or due to \VT{}'s dynamicity). 
In this paper, we build on the insights in ~\cite{hurier2016lack,mohaisen2014av,mohaisen2013towards} to highlight the aspects of \VT{}'s dynamicity that impact the verdicts given by scanners and how this impacts the performance of threshold-based labeling strategies and, in turn, the reliability of malware detection methods built on top of their labels.

\textbf{Accurate Labeling Strategies.}
Aware of their sensitivity to \VT{}'s dynamicity, researchers have attempted to replace threshold-based labeling strategies with more sophisticated labeling strategies, primarily based on \gls{ml}. 
In ~\cite{kantchelian2015better}, Kantchelian et al.\ used the \VT{} scan reports of around 280K binaries to build two \gls{ml}-based techniques to aggregate the results of multiple scanners into a single ground-truth label for every binary. 
In the first technique, Kantchelian et al.\ assume that the ground truth of an app (i.e., malicious or benign), is unknown or \emph{hidden}, making the problem of estimating this ground truth is that of unsupervised learning. 
Furthermore, they assumed that the verdicts of more consistent, less erratic scanners are more likely to be correlated with the correct, hidden ground truth than more erratic scanners. 
Thus, more consistent scanners should have larger weights associated with their verdicts. 
To estimate those weights and, hence, devise an unsupervised \gls{ml}-based labeling strategy, the authors used an \gls{em} algorithm based on a Bayesian model to estimate those models. 
The second technique devised by Kantchelian et al.\ is a supervised one based on regularized logistic regression. 
However, the authors did not describe the nature of the features they use to train such an algorithm. 
To devise an automated method to label apps based on different verdicts given by antiviral scanners, Sachdeva et al.\ \cite{sachdeva2018android} performed measurements to determine the most correct \VT{} scanners using scan reports of a total of 5K malicious and benign apps.
Using this information, they assign a weight to each scanner that they use to calculate a malignancy score for apps based on their \VT{} scan reports. 
Depending on manually-defined thresholds, the authors use this score to assign a confidence level of Safe, Suspicious, or Highly Suspicious to test apps. 
The works in ~\cite{kantchelian2015better} and ~\cite{sachdeva2018android} seem to dismiss threshold-based labeling strategies and go for sophisticated labeling methods that may be difficult to comprehend and utilize. 
In our paper, we studied the reasons behind the fluctuating performance of this type of labeling strategies and found that, despite their simplicity, they can be effectively utilized to accurately label Android apps based on their \VT{} scan reports. 
Furthermore, we provide the research community with an algorithm to workaround \VT{}'s dynamicity and yield the current optimal thresholds that would yield accurate labels. 
\section{Conclusion}
\label{sec:conclusion}
The infeasibility of manually analyzing and labeling Android apps and the lack of more stable alternatives forces the research community to use the online platform, \VT{}, to label apps they use in training and evaluating malware detection methods. 
Although \VT{} is known to be dynamic and volatile, previous research neither delved into the aspects of the platform's dynamicity, how it impacts the threshold-based labeling process, and how to work around it nor implemented alternatives to \VT{}. 

With a focus on Android apps, in this paper, we studied \VT{} to identify how its alleged dynamicity manifests itself and how it impacts threshold-based labeling strategies that are widely-adopted within the research community. 
Using our findings, we provided a method that bypasses the aspects of \VT{}'s dynamicity to find the thresholds of scanners that would yield the best labeling accuracies and more effective and reliable malware detection methods. 
Despite this method, we realize that \VT{}'s limitations ultimately calls for the replacement of the online platform with more reliable alternatives. 
In order not to implement alternative platforms that suffer from the same shortcomings of \VT{}, we discussed the four limitations of \VT{} that we identified through our measurements and analysis and how to avoid them in building alternative platforms.

\bibliographystyle{IEEEtran}
\bibliography{virustotal}

\begin{thebibliography}{10}
\providecommand{\url}[1]{#1}
\csname url@samestyle\endcsname
\providecommand{\newblock}{\relax}
\providecommand{\bibinfo}[2]{#2}
\providecommand{\BIBentrySTDinterwordspacing}{\spaceskip=0pt\relax}
\providecommand{\BIBentryALTinterwordstretchfactor}{4}
\providecommand{\BIBentryALTinterwordspacing}{\spaceskip=\fontdimen2\font plus
\BIBentryALTinterwordstretchfactor\fontdimen3\font minus
  \fontdimen4\font\relax}
\providecommand{\BIBforeignlanguage}[2]{{%
\expandafter\ifx\csname l@#1\endcsname\relax
\typeout{** WARNING: IEEEtran.bst: No hyphenation pattern has been}%
\typeout{** loaded for the language `#1'. Using the pattern for}%
\typeout{** the default language instead.}%
\else
\language=\csname l@#1\endcsname
\fi
#2}}
\providecommand{\BIBdecl}{\relax}
\BIBdecl

\bibitem{hurier2017euphony}
M.~Hurier, G.~Suarez-Tangil, S.~K. Dash, T.~F. Bissyand{\'e}, Y.~L. Traon,
  J.~Klein, and L.~Cavallaro, ``Euphony: Harmonious unification of cacophonous
  anti-virus vendor labels for android malware,'' in Proceedings of the 14th
  International Conference on Mining Software Repositories.\hskip 1em plus
  0.5em minus 0.4em\relax IEEE Press, 2017, pp. 425--435.

\bibitem{hurier2016lack}
M.~Hurier, K.~Allix, T.~F. Bissyand{\'e}, J.~Klein, and Y.~Le~Traon, ``On the
  lack of consensus in anti-virus decisions: Metrics and insights on building
  ground truths of android malware,'' in International Conference on Detection
  of Intrusions and Malware, and Vulnerability Assessment.\hskip 1em plus 0.5em
  minus 0.4em\relax Springer, 2016, pp. 142--162.

\bibitem{miller2016reviewer}
B.~Miller, A.~Kantchelian, M.~C. Tschantz, S.~Afroz, R.~Bachwani,
  R.~Faizullabhoy, L.~Huang, V.~Shankar, T.~Wu, G.~Yiu et~al., ``Reviewer
  integration and performance measurement for malware detection,'' in
  International Conference on Detection of Intrusions and Malware, and
  Vulnerability Assessment.\hskip 1em plus 0.5em minus 0.4em\relax Springer,
  2016, pp. 122--141.

\bibitem{kantchelian2015better}
A.~Kantchelian, M.~C. Tschantz, S.~Afroz, B.~Miller, V.~Shankar, R.~Bachwani,
  A.~D. Joseph, and J.~D. Tygar, ``Better malware ground truth: Techniques for
  weighting anti-virus vendor labels,'' in Proceedings of the 8th ACM Workshop
  on Artificial Intelligence and Security.\hskip 1em plus 0.5em minus
  0.4em\relax ACM, 2015, pp. 45--56.

\bibitem{virustotal2019}
\BIBentryALTinterwordspacing
VirusTotal. Virustotal. [Online]. Available: \url{http://tiny.cc/xjbb7y} (2019)
\BIBentrySTDinterwordspacing

\bibitem{wang2018beyond}
H.~Wang, Z.~Liu, J.~Liang, N.~Vallina-Rodriguez, Y.~Guo, L.~Li, J.~Tapiador,
  J.~Cao, and G.~Xu, ``Beyond google play: A large-scale comparative study of
  chinese android app markets,'' in Proceedings of the Internet Measurement
  Conference 2018.\hskip 1em plus 0.5em minus 0.4em\relax ACM, 2018, pp.
  293--307.

\bibitem{li2017understanding}
L.~Li, D.~Li, T.~F. Bissyand{\'e}, J.~Klein, Y.~Le~Traon, D.~Lo, and
  L.~Cavallaro, ``Understanding android app piggybacking: A systematic study of
  malicious code grafting,'' IEEE Transactions on Information Forensics and
  Security, vol.~12, no.~6, 2017, pp. 1269--1284.

\bibitem{suarez2017droidsieve}
G.~Suarez-Tangil, S.~K. Dash, M.~Ahmadi, J.~Kinder, G.~Giacinto, and
  L.~Cavallaro, ``Droidsieve: Fast and accurate classification of obfuscated
  android malware,'' in Proceedings of the Seventh ACM on Conference on Data
  and Application Security and Privacy.\hskip 1em plus 0.5em minus 0.4em\relax
  ACM, 2017, pp. 309--320.

\bibitem{wei2017deep}
F.~Wei, Y.~Li, S.~Roy, X.~Ou, and W.~Zhou, ``Deep ground truth analysis of
  current android malware,'' in International Conference on Detection of
  Intrusions and Malware, and Vulnerability Assessment.\hskip 1em plus 0.5em
  minus 0.4em\relax Springer, 2017, pp. 252--276.

\bibitem{yang2017malware}
W.~Yang, D.~Kong, T.~Xie, and C.~A. Gunter, ``Malware detection in adversarial
  settings: Exploiting feature evolutions and confusions in android apps,'' in
  Proceedings of the 33rd Annual Computer Security Applications
  Conference.\hskip 1em plus 0.5em minus 0.4em\relax ACM, 2017, pp. 288--302.

\bibitem{pendlebury2019}
R.~J. J.~K. Feargus~Pendlebury, Fabio~Pierazzi and L.~Cavallaro, ``Tesseract:
  Eliminating experimental bias in malware classification across space and
  time,'' in 28th USENIX Security Symposium.\hskip 1em plus 0.5em minus
  0.4em\relax Santa Clara, CA: USENIX Association, 2019.

\bibitem{peng2019opening}
P.~Peng, L.~Yang, L.~Song, and G.~Wang, ``Opening the blackbox of virustotal:
  Analyzing online phishing scan engines,'' in Proceedings of the Internet
  Measurement Conference.\hskip 1em plus 0.5em minus 0.4em\relax ACM, 2019, pp.
  478--485.

\bibitem{mohaisen2014av}
A.~Mohaisen and O.~Alrawi, ``Av-meter: An evaluation of antivirus scans and
  labels,'' in International Conference on Detection of Intrusions and Malware,
  and Vulnerability Assessment.\hskip 1em plus 0.5em minus 0.4em\relax
  Springer, 2014, pp. 112--131.

\bibitem{salem2018poking}
A.~Salem and A.~Pretschner, ``Poking the bear: Lessons learned from probing
  three android malware datasets,'' in Proceedings of the 1st International
  Workshop on Advances in Mobile App Analysis.\hskip 1em plus 0.5em minus
  0.4em\relax ACM, 2018, pp. 19--24.

\bibitem{pendlebury2018enabling}
F.~Pendlebury, F.~Pierazzi, R.~Jordaney, J.~Kinder, and L.~Cavallaro,
  ``Enabling fair ml evaluations for security,'' in Proceedings of the 2018 ACM
  SIGSAC Conference on Computer and Communications Security.\hskip 1em plus
  0.5em minus 0.4em\relax ACM, 2018, pp. 2264--2266.

\bibitem{sanders2017garbage}
H.~Sanders and J.~Saxe, ``Garbage in, garbage out: how purportedly great ml
  models can be screwed up by bad data,'' Technical report, 2017.

\bibitem{tam2017evolution}
K.~Tam, A.~Feizollah, N.~B. Anuar, R.~Salleh, and L.~Cavallaro, ``The evolution
  of android malware and android analysis techniques,'' ACM Computing Surveys
  (CSUR), vol.~49, no.~4, 2017, p.~76.

\bibitem{arshad2016android}
S.~Arshad, M.~A. Shah, A.~Khan, and M.~Ahmed, ``Android malware detection \&
  protection: a survey,'' International Journal of Advanced Computer Science
  and Applications, vol.~7, 2016, pp. 463--475.

\bibitem{allix2016androzoo}
K.~Allix, T.~F. Bissyand{\'e}, J.~Klein, and Y.~Le~Traon, ``Androzoo:
  Collecting millions of android apps for the research community,'' in Mining
  Software Repositories (MSR), 2016 IEEE/ACM 13th Working Conference on.\hskip
  1em plus 0.5em minus 0.4em\relax IEEE, 2016, pp. 468--471.

\bibitem{arp2014drebin}
D.~Arp, M.~Spreitzenbarth, M.~Hubner, H.~Gascon, and K.~Rieck, ``Drebin:
  Effective and explainable detection of android malware in your pocket.'' in
  NDSS, 2014.

\bibitem{sklearn2019mcc}
\BIBentryALTinterwordspacing
scikit learn. sklearn.metrics.matthews\_corrcoef. [Online]. Available:
  \url{http://tiny.cc/8gbb7y} (2019)
\BIBentrySTDinterwordspacing

\bibitem{sklearn2019accuracy}
\BIBentryALTinterwordspacing
------. sklearn.metrics.matthews\_corrcoef. [Online]. Available:
  \url{http://tiny.cc/x65xlz} (2019)
\BIBentrySTDinterwordspacing

\bibitem{appinventor2019}
\BIBentryALTinterwordspacing
M.~A. Inventor. About us - explore mit app inventor. [Online]. Available:
  \url{http://tiny.cc/byp89y} (2019)
\BIBentrySTDinterwordspacing

\bibitem{avtest2019}
\BIBentryALTinterwordspacing
A.-T. T. I. I.-S. Institute. The best antivirus software for android. [Online].
  Available: \url{http://tiny.cc/1k66az} (2019)
\BIBentrySTDinterwordspacing

\bibitem{pcmagazin2008bitdefender}
\BIBentryALTinterwordspacing
P.~Magazin. Bitdefender free edition. [Online]. Available:
  \url{http://tiny.cc/8vx9az} (2008)
\BIBentrySTDinterwordspacing

\bibitem{androguard2018}
\BIBentryALTinterwordspacing
androguard. androguard: Reverse engineering, malware and goodware analysis of
  android applications ... and more (ninja !). [Online]. Available:
  \url{https://goo.gl/toC9dB} (2018)
\BIBentrySTDinterwordspacing

\bibitem{salem2019towards}
J.~N. Aleieldin~Salem, Michael~Hesse and A.~Pretschner, ``Towards empirically
  assessing behavior stimulation approaches for android malware,'' in The 13th
  International Conference on Emerging Security Information, Systems and
  Technologies.\hskip 1em plus 0.5em minus 0.4em\relax International Academy,
  Research and Industry Association (IARIA), 2019.

\bibitem{k9mail2019}
\BIBentryALTinterwordspacing
K-9 mail - advanced email for android. [Online]. Available:
  \url{http://tiny.cc/1fbb7y} (2019)
\BIBentrySTDinterwordspacing

\bibitem{mohaisen2013towards}
A.~Mohaisen, O.~Alrawi, M.~Larson, and D.~McPherson, ``Towards a methodical
  evaluation of antivirus scans and labels,'' in International Workshop on
  Information Security Applications.\hskip 1em plus 0.5em minus 0.4em\relax
  Springer, 2013, pp. 231--241.

\bibitem{sachdeva2018android}
S.~Sachdeva, R.~Jolivot, and W.~Choensawat, ``Android malware classification
  based on mobile security framework,'' IAENG International Journal of Computer
  Science, vol.~45, no.~4, 2018.

\end{thebibliography}

\end{document}